\documentclass[aip,amsmath,amssymb,reprint]{revtex4-1}
\pdfoutput=1
\usepackage{graphicx}
\usepackage{caption}
\usepackage{subcaption}
\usepackage{mathpazo}
\usepackage{dcolumn}
\usepackage{bm}

\usepackage[utf8]{inputenc}
\usepackage[T1]{fontenc}
\usepackage{fonttable}
\usepackage{mathptmx}
\usepackage{color, soul}
\begin{document}
\preprint{AIP/123-QED}

\title{Simple, Reversible Gradient Seebeck Coefficient Measurement System for 300 - 600K}

\author{Soumya Biswas}
\thanks{These two authors contributed equally}

\author{Aditya S Dutt}
\thanks{These two authors contributed equally}
\author{Nirmal Sabastian}%

\author{Vinayak B Kamble}
\email{kbvinayak@iisertvm.ac.in}
\affiliation{School of Physics, Indian Institute of Science Education and Research, Thiruvananthapuram, Kerala, India 695551}%
\date{\today}

\begin{abstract}
An in-house Seebeck coefficient measurement system has been developed which can measure the thermoemf (Seebeck coefficient) of the sample, under large temperature difference, in the temperature range 300-600 K. Unlike majority of reported instrumental designs, the system does not have a hot walled chamber and hence is much closer to real time thermoelectric applications conditions. The system consists of two brass blocks supported heaters. These heaters are placed on either side of the sample through silver caps, thus allows individual temprature control . A reversible temperature gradient is applied across the sample and the measurement is carried out in quasi-static direct current mode. Hence, a more accurate Seebeck coefficient measurement is obtained. By virtue of its design the sample holder ensures a minimum thermal and electrical contact resistance during a measurement cycle. The combination of metals used for measurement (Ag and Cu) shows negligible junction contribution. The variance upto $\pm$ 2\% and accuracy upto 7\% at high temperature has been obtained using calibration samples' reference data of state of the art commercial system.
\end{abstract}

\maketitle

\section{\label{sec:level1}Introduction}

Thermoelectric effect is a phenomenon by which a temperature gradient generates an electric potential or vice versa\cite{tritt2006, sootsman2009, Mahan2016}. It does have a great relevance now, since we are overly depending on non-renewable sources of energy, which shall not be replenished over time. Thus, by using an engine having an added thermoelectric power generation unit, it can convert this dissipated energy into usable electrical energy, which in turn increases the efficiency of the engine. However, in order to tap its complete potential, a high efficiency thermoelectric material is highly desirable for a practical applications\cite{sootsman2009, he2017}.
 The “Figure of merit” is the quantity, which represents the efficiency of a thermoelectric material. It is represented as "zT" and the expression for the same is as shown by Eq. (1)\cite{Mahan2016, borup2015}

\begin{equation}
 zT = \frac{S^2 \sigma T}{\kappa}
\end{equation}

Where, \\
S is the thermopower (also known as Seebeck coefficient); \\
$\sigma$  is the electrical conductivity; \\
$\kappa$ is the total thermal conductivity and \\
T is the absolute temperature. \\
The quantity in the numerator ${S^2}\sigma$, is a measure of power output and is called as power factor.
Thus, for the estimation of thermoelectric efficiency, one has to measure Seebeck coefficient, electrical conductivity and thermal conductivity of the given material.\cite{borup2015} Although, there are several of commercial systems available for the Seebeck measurement in diverse temperature range, they are fairly expensive and need a periodic maintenance to keep them in working condition. Besides, thye cannot be customized as per the experimental requirements such as with magnetic field or gas ambience. Hence, we have developed a very low cost system (about 20\(\%\) of the cost of the commercial system), which offers similar accuracy as that of the prior in a limited temperature range. Thus, the objective of this study is to design, fabricate and calibrate a setup that can measure the Seebeck coefficient of the sample in a given temperature range (300 - 650 K), which is good enough for study of a range of class of compounds such as chalcogenides, oxides and other alloys.
A number of designs have been reported in literature till date.\cite{wood1988, singh2017, mishra2015, borup2015, nakama1998, Titas2005, Zhu2017, martin2010, iwanaga2011, gunes2014, Kumar2019} However, many of them are quite complex or involve a hot zone furnace\cite{ponnambalam2006, Titas2005} etc increasing the cost of the instrument as well as the power requirements. Moreover, such designs require high temperature stable refractory materials such as Alumina and Platinum for electrically conducting leads. Here, because of the smaller size of the heaters, the hot zone does not extend much further away from the sample and does not impose stringent high temperature stability requirements for many of the components. Further, the entire system is enclosed in stainless steel chamber which allows to control the environment.

\begin{figure*}[t]
\includegraphics[width=1\textwidth]{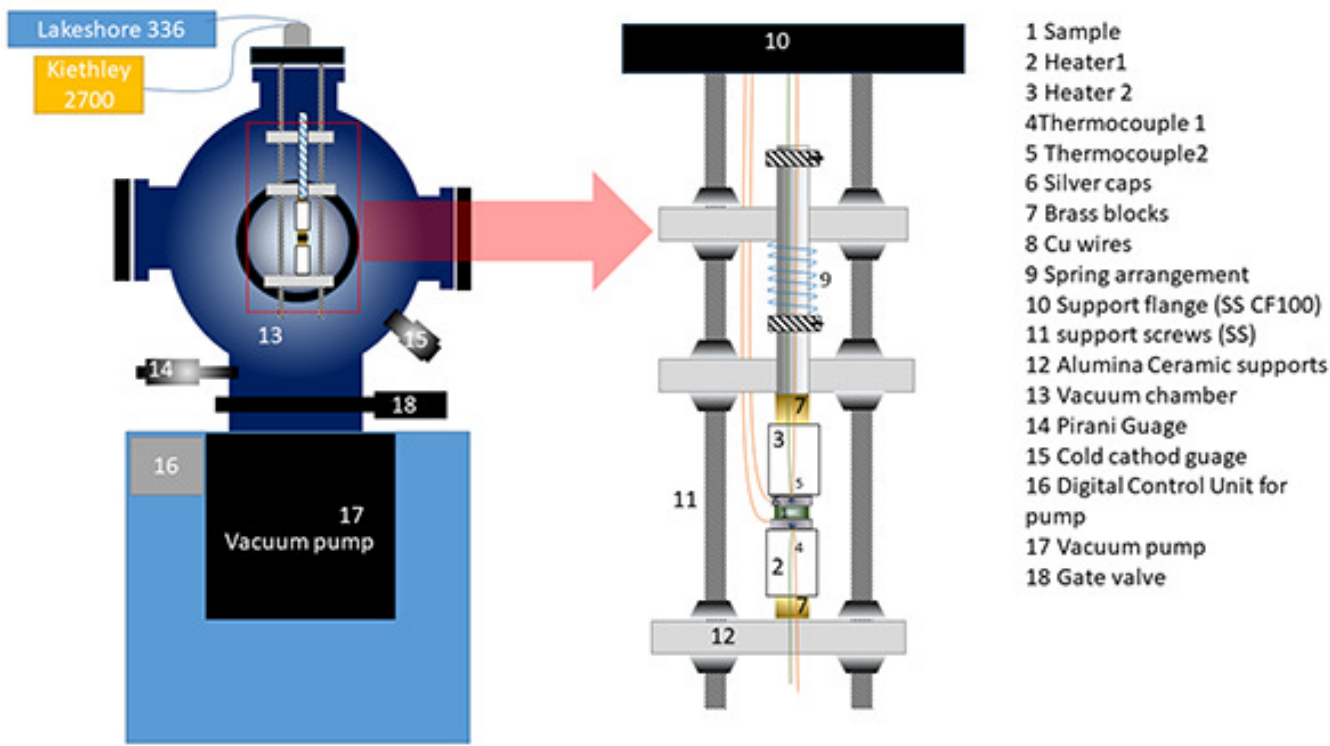}
\caption{(a) The schematic diagram of the Seebeck set-up with enlarged view of the assembly.}
\label{Fig1}
\end{figure*}

\begin{figure}[b]
\includegraphics[width=0.45\textwidth]{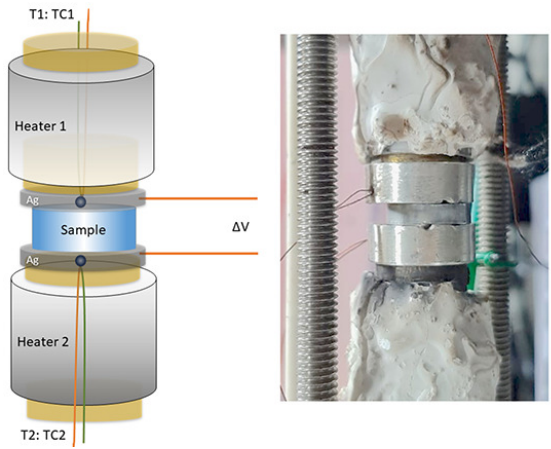}
\caption{A schematic diagram and a digital photograph showing thermocouple and voltage lead junctions in contact with the sample.}
\label{Fig2}
\end{figure}

\section{\label{sec:level1}Materials and Methods}
\subsection{\label{sec:level2}Design of the Apparatus}

Fig. \ref{Fig1} shows the schematic design of the system for measuring the Seebeck co-efficient of a sample. A home built system is designed which can measure the thermoemf by applying a small temperature difference across the sample. Here, the sample is sandwiched between two heaters, which are placed on two solid brass cylinders as shown in the Fig. \ref{Fig1}.
It has taken a significant effort to optimize the design of the setup. The system is calibrated and interfaced for data acquisition system described below. The system is designed to work for temperature range of RT - 350 $^{o}$C, which is ideal for most chalcogenide and a large family of oxides and alloys.

\subsection{\label{sec:level2}Control and Data Acquisition System}
Here, the sample (blue colored object in the schematic diagram shown in Fig. \ref{Fig2}) is sandwiched between a pair of brass cylinders having silver cap electrodes. Heating wires (nichrome) are wound on the brass cylinders and followed by heater insulation in order to minimize the heat loss and allowing greater control over temperature. These heaters provide the temperature difference across the sample, which are measured as well as controlled using two “K” type thermocouples inserted inside these brass cylinders centrally as seen in figure. The thermocouples are electrically insulated from the brass rods using mica sheets. While, the voltage produced, is measured between two copper wires drawn from either silver caps attached to each brass rod. Silver was chosen for its high electrical ($6.25 \times 10^{5} \Omega^{-1}.m$) and thermal conductivity (406 W/K.m)\cite{kharote2017}. The assemblly is held with the help of two long SS screws anchored form a CF100 flnage. All the wires are soldered to a circular connector at the center of the mounting flange allowing complete enclosure. One of the brass rod is mounted on to a spring loaded arrangement as shown in Fig. \ref{Fig1}. The entire assembly is inserted into a a vacuum chamber connected to a vacuum pump, allowing control over the environment i.e. either in ambient or ultra high vacuum (10$^{-5}$ torr).

The heaters were controlled using Lakeshore 336 temperature controller having four control outputs and two thermocouple controls. Two K type thermocouples (30 guage) were connected to the Lakeshore temperature controller.
A temperature difference of about $\pm 10$ $^o$C is given across sample reversibly. As a result of this temperature difference, a small voltage (usually a few mili volts) is produced, which is measured with the help of a keithley 2700 digital multimeter having a 7700 multiplexure card of 20 channels. 

\begin{figure}[t]
\includegraphics[width=0.45\textwidth]{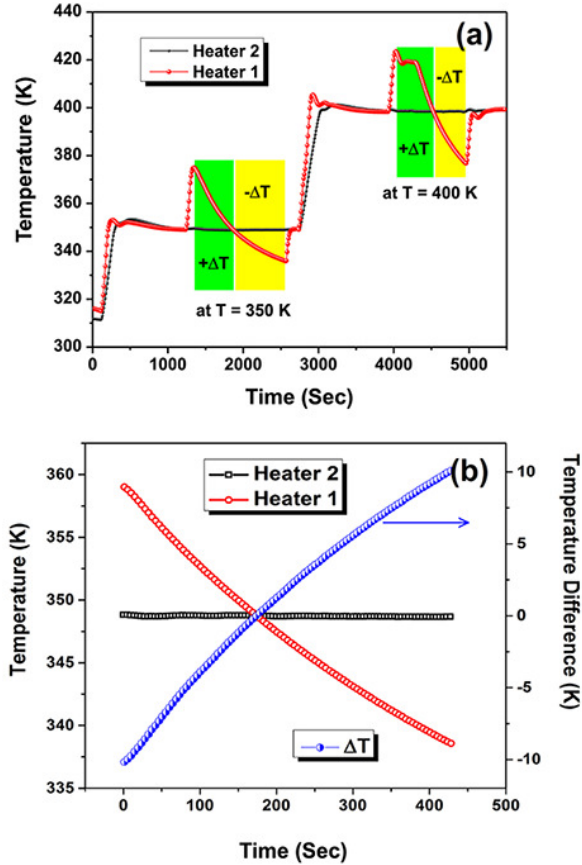}
\caption{(a) The typical measurement procedure of Seebeck measurement at two different temperatures i.e. 350 and 400 K. (b) The variation of temperature profile of the two heaters across the sample with temperature difference}
\label{test}
\end{figure}

\section{Results}
Two K type thermocouples are placed in either silver caps through insulation, to measure the temperature. Owing to high thermal conductivity of silver, it is assumed that the sample and silver cap's flat surface are in thermal equilibrium. One of the block is attached to a spring, so that it allows easy placement and removal of sample, at the same time, ensure a good electrical and thermal contact between the metal blocks and the sample thorugh pressure. The K type thermocouples are composed of alumel and chromel junction which has its own seebeck of 40 $\mu$V/K and which is linear across the entire measurement range.

The copper wires were inserted into a hole in the silver cap, which served as voltage leads. The voltage leads and the thermocouples are connected to the digital multimeter. The 7700 scanner card has 20 channels allowing to do simultaneous measurements of 20 physical quantities such as resistance, temperature, voltage, current etc. 

\begin{figure}[t]
\includegraphics[width=0.45\textwidth]{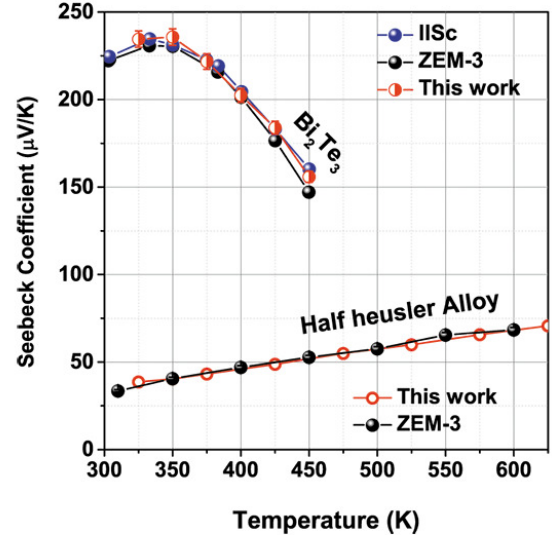}
\caption{The comparison of the seebeck data of Bi$_2$Te$_3$ and Half heusler alloy with reference data.}
\label{Fig4}
\end{figure}

\subsection{Measurement procedures}
In order to use the above configuration for measuring the voltage, it is mandatory to confirm that all the junctions are ohmic in nature i.e. they allow reversible charge flow. A set of I-V measurement, were taken at each of the stage to ensure the proper electrical contact. 
The measurement is carried out in quasi-steady state differnetial fasion. This method is time efficient and practical.\citep{martin2010} The Fig \ref{test}(a) shows the profile of temperatures across the samples for both the heaters. This demonstrates either polarities of the temperature gradient and hence more reliable data. The chamber was evacuated using rotary pump and then pressurized to near atmospheric value (760 Torr) using dry nitrogen gas before beginning the measurement. The sample was brought to measurement temperature by heating form either side to a desired value (for instance at 350 K as shown in Fig 3(a)). Subsequently, a small temperature difference is applied. In a typical measurement the heater 2 is held at a constant measurement temperature (for instance T K) and the temperature of heater 1 is varied from T + 10 K. The Heater 1 is then set to T - 10K from Heater 2. This allows the sample to cool naturally. The data of voltage produced is recorded during cooling ensuring a steady state measurement. The typical measurement interval of the system  i.e the time interval between measurement of temperature difference and the voltage, is about 100-150 mS. However, most of hot walled measurement systems like ZEM-3, first heats the sample to a desired temperature either by resistive heating or Infra Red radiation and then give a small temperature difference on one side. The major drawback of commercial system design is, it needs a tall sample of at least 6 mm in height for accurate measurements. Whereas, here the data of $Bi_2Te_3$ reported has been performed on sample of 2 mm thickness. The horizontal dimensions of the sample in our design can be anywhere between 3 to 12 mm.\\
The two samples chosen for calibration have very different S value (by an order of magnitude) as well as the trend with temperature. The data so obtained are compared with those measured on state of the art systems from reputed laboratories working in thermoelectrics from the country. The calibration was done with two samples of known Seebeck value (p-type bismuth telluride and half heusler alloy sample) from standard systems as shown in Fig \ref{Fig4}. It may be seen the values obtained are found to be consistent with the commercial ZEM-3 system. 

\subsection{Error estmation}
\begin{figure}[]
\includegraphics[width=0.45\textwidth]{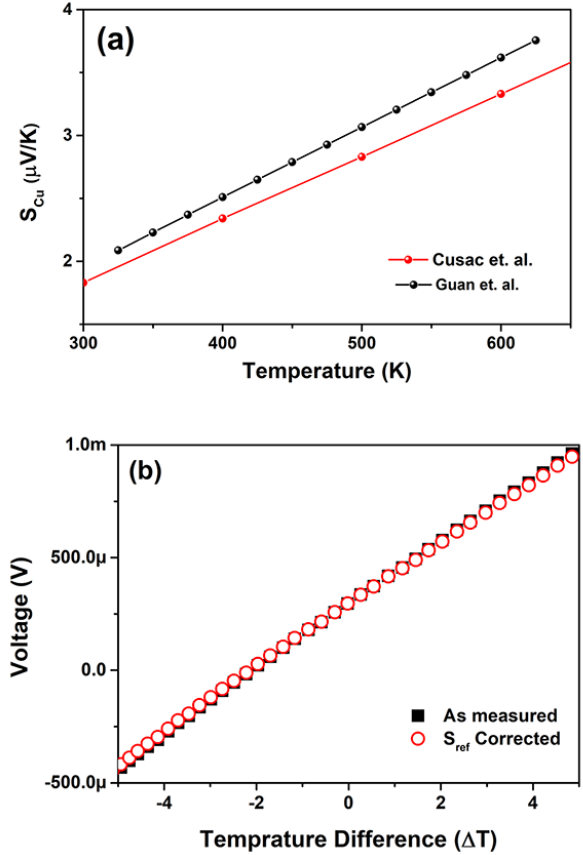}
\caption{(a) The Seebeck Coefficient values of Cu as a  function of temperature from literature\cite{cusack1958, guan2013} and (b) The comparison of voltage measured before and after correctiong the wire seebeck using eq \ref{Eff_S_formula} at 500K.}
\label{Fig6}
\end{figure}

The voltage produced between two leads of the voltmeter can be expressed as
\begin{equation}
-\Delta V = \int_{Term1}^{Term2} E.dl\\
 = \int_{Term1}^{Term2} S(T). dT.
\end{equation}
Where, E is the electric field developed over length dl due to thermoemf (S).

This may be written as-
\begin{equation}
\begin{aligned}
-\Delta V = \int_{T_a}^{T_1} S_{Cu} dT + \int_{T_1}^{T_1} S_{Ag} dT  +  \int_{T_1}^{T_2} S(T).dT \\* + \int_{T_2}^{T_2} S_{Ag} dT  + \int_{T_2}^{T_b} S_{Cu} dT
\end{aligned}
\end{equation}

Here, $T_2- T_1 \ll T$, and $T_a = T_b $ (The temperature of connector)
Besides, the temperature within the silver blocks does not change and $S_{Ag}$  also does not vary significantly over $ \Delta T$ then we can write \cite{mackey2014, de2013},
\begin{equation}\label{Eff_S_formula}
-\Delta V =  \int_{T_1}^{T_2} S(T) dT  + \int_{T_1}^{T_2} S_{wire} dT
\end{equation} 
where
\begin{equation}
 \int_{T_1}^{T_2} S_{wire}  dT = \int_{T_b}^{T_2} S_{Cu} dT' - \int_{T_1}^{T_b} S_{Cu} dT''  
\end{equation}
Since, dT' and dT" are the temperature difference between $T_b$ and $T_2$ and $T_b$ and $T_1$ respectivelys such that,
 \begin{equation}
 dT = |dT'' - dT'|  
\end{equation}
Thus, the uncertainly in the measured voltage and actual sample voltage is evaluated by correcting with the seebeck coefficient of the wire material i.e. Cu at measurment temperature and the corrosponding temperature difference. The values of seebeck coefficient have been obtained from literature \cite{cusack1958, guan2013} and plotted at shown in Fig \ref{Fig6}. Here, the advantage of using steady state differential method is evident that it does not imposes the  requirement of curves intersecting the ordant, i.e V=0 at $\Delta T$ =0. This offset might arise from thermocouple inhomogenities and possibility of non homogenous contact interfaces.\citep{martin2010}. However, the presence of offset does not affect the slope of the data and hence is ignored. (it is also found in data measured using Commercial Zem-3 system at high temperatures\citep{mackey2014}.

Thus, the close agreement in values obtained between the system developed and that of the commercial system signifies the goodness of the data obtained and hence can be used for characterizing new samples.
\begin{figure}[b]
\includegraphics[width=0.45\textwidth]{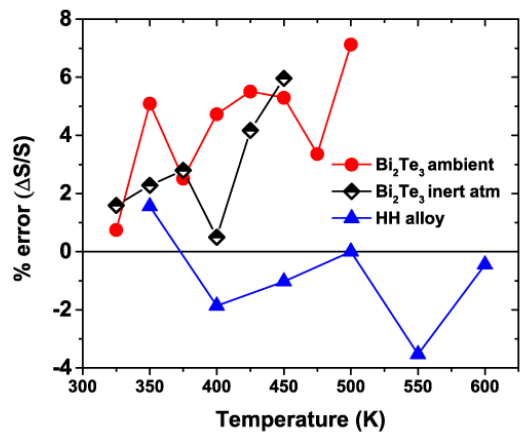}
\caption{The estimated error of the seebeck coefficient of Bi$_2$Te$_3$ and Half Heusler (HH) alloy ($Zr_{0.75}Ti_{0.25}NiSn_{0.97}Si_{0.03}$) with reference data.}
\label{Fig5}
\end{figure}

The statistical varition in the data shown, is normally same as the size of the legend in Fig \ref{Fig4}, however, is barely seen in case Bi$_2$Te$_3$ due to high seebeck value. This is evaluated by statistical mean of several measurements of same sample at same temperatures and it has been observed that the variance of data obtained is upto $\pm 2\%$ of the full scale. Further, the important parameter of error has been evaluated by estimating the difference between measured average value and ther reference value ($\Delta S$) normalized with the reference value (S). This percentage error has been shown in Fig \ref{Fig5} as a function of sample temperature. It may be noted that the value is max $\pm 7 \%$ at high temperatures, which is similar to uncertainty in ZEM-3 system \cite{mackey2014} where it arises predominently due to cold finger effect of thermocouples and electrical leads.

\subsection{Optimization of system}
The system has undergone several changes in order to optimize the data accuracy. Some were empirical, while many were leart from literature like Martin et al. and others \citep{martin2010, de2013, iwanaga2011}. Here, the following changes were made for optimzations.
\begin{itemize}
\item The copper blocks were used intially which were found to corrode/oxidize at high temperature in ambient atmosphere. Hence, brass blocks were used which shows higher melting temprature, thus better thermal as well as chemical stability. Besides, brass are softer and hence easily machinable. 
\item However, the junction of brass block and copper wire was found to introduce significant voltage and hence the silver cap was used which acted as electrode and also has same seebeck value as that of copper\cite{mackey2014, de2013}.
\item When brass blocks were longer than the heater area, there used to instability in the tempearature due to continuous cooling of exposed brass surface. Hence, the same was covered with thermal insulation for retaining the heat, which also improved temperature stability.
\item When thermocouples are not electrically insulated from the metal block in which they are inserted, it leads to discrepency in the thermocouple voltage due to another junction. Hence, insulating thermocouples with high thermal conductivity material like mica avoids this spurious error. 
\item Besides, some samples like $Bi_2Te_3$ and oxides shows a change in seebeck value when meausured in oxygen atmosphere (i.e. ambient) and inert atmosphere. Hence, introducing the controlled atmosphere has been beneficial for more accurate measurement. However, some robust samples like the half heusler alloy ($Zr_{0.75}Ti_{0.25}NiSn_{0.97}Si_{0.03}$) measured in this study did not show noticable change when measured in ambient or inert atmosphere. 
\end{itemize}

\section{Summary}
The simple, low cost, Seebeck co-efficient measurement system has been made in-house and calibrated at higher temperature (300 - 600 K) with a reference samples measured in the commercial machine. The sample values were in good agreement with the commercial Seebeck ZEM-3 system by ULVAC RIKO. The measurement were conducted under sufficiently large, reversible temperature gradient ( about $\pm$5-8 K). This demonstrates the utility of the system with minimal economical investement while giving the similar accuracy of the data ($\pm 7 \%$ ).

\section{Note}
The current affiliations of AD is TU Dresden and that of NS is IISc Bangalore. The work was performed during their affiliation at IISER Thiruvananthapuram.

\begin{acknowledgments}
The authors would like to thank Prof. Satish Vitta from MEMS, IIT Bombay for providing the half heusler alloy sample and its data measured on ULVAC Riko system. Moreover, we are also thankful to Prof. Arun Umarji,  Mr Rajasekhar of IISc Bangalore for the Bi$_2$Te$_3$ sample and its data.
\end{acknowledgments}
\bibliographystyle{aip}

\end{document}